\documentclass[twocolumn,pra,showpacs,floatfix,preprintnumbers,amsmath,amssymb]{revtex4}
\usepackage{dcolumn}
\usepackage{graphicx}
\usepackage{bm}

\begin{document}

\title{Compact solid-state laser source for $1S-2S$ spectroscopy in atomic hydrogen }

\author{N. Kolachevsky}
\altaffiliation{P.N. Lebedev Physics
Institute, Leninsky prosp. 53, 119991 Moscow, Russia}
\author{J. Alnis}
\altaffiliation{University of Latvia, 19 Ranis Blvd., LV-1586 Riga,
Latvia}
\author{S. D. Bergeson}
\altaffiliation{Brigham Young University, N283 ESC, Provo, Utah 84602, USA}
\author{T. W. H\"{a}nsch}
\altaffiliation{Ludwig--Maximilians--University,
Munich, Germany}

\affiliation{Max--Planck--Institut f\"{u}r
Quantenoptik, Hans--Kopfermann--Str. 1, 85748 Garching,
Germany}

\date{\today}

\begin{abstract}
We demonstrate a novel compact solid-state 
laser source  for high-resolution
two-photon spectroscopy of the $1S-2S$ transition in atomic
hydrogen. The source emits up to 20 mW  at 243 nm and
consists of a 972 nm diode laser, a tapered amplifier, and two
doubling stages. The diode laser is actively stabilized to a
high-finesse cavity. We compare the new source to the stable 486
nm dye laser used in previous experiments and
record $1S-2S$ spectra using both systems. With the solid-state
laser system we demonstrate a resolution of the hydrogen
spectrometer of $6\times 10^{11}$ which is promising for a number of
high-precision measurements in hydrogen-like systems.
\end{abstract}


\maketitle

Since the first experiments in the late 1970´s \cite{Haensch1},
$1S-2S$ spectroscopy in atomic hydrogen and deuterium has
provided essential data for fundamental physics.
These include the determination of the Rydberg constant and tests of
quantum electrodynamics theory \cite{Biraben, Weitz2s4s},
determination of nuclear properties of the proton and deuteron
\cite{Weitz2s4s, Huberisotop}, spectroscopy of Bose-Einstein
condensate in hydrogen \cite{Kleppner}, hyperfine structure
measurements of the $2S$ state \cite{KolPRA}, and searching for
the drift of the fine structure constant
\cite{Fischetal}.

In all of these experiments, the $2S$ state was
excited by the second harmonic of a stabilized dye laser
operating at 486 nm. The dye laser system originally developed in
Garching has been repeatedly upgraded to meet the stringent
requirements of high-resolution spectroscopic experiments.
The uncertainty in recent $1S-2S$ frequency measurements in
atomic hydrogen is 25 Hz ($\Delta f / f_0 = 1.0\times 10^{-14}$)
\cite{Fischetal}.  However, an uncertainty limit
near the natural linewidth of 1.3 Hz is desirable.
In addition to improving the accuracy of these measurements, it
would be helpful to reduce the size of the experiment.
This is not only necessary for
metrological applications such as creating a transportable optical
frequency reference, but also for opening possibilities for completely
new experiments to test some fundamental aspects of physics.

For example, a number of recently proposed experiments concern
spectroscopy of exotic atomic systems which are not available in
the laboratory.
This will require a new laser system to be set up close to the
place where the atomic sample is produced.
Experiments to study the $1S-2S$ spectroscopy in anti-hydrogen
are prepared
by the collaborations ATHENA (ALPHA) \cite{Athena} and ATRAP
\cite{Atrap} at CERN. A comparison between hydrogen and
anti-hydrogen spectra should provide one of the most stringent
tests of the CPT theorem. A frequency measurement of the  $1S-2S$
transition in tritium  could provide new independent information
on the triton charge radius and its polarizability \cite{86GI2A}.
In addition, there is an on-going activity on optical spectroscopy
of positronium and muonium \cite{positronium, Jungmann, Habs}.

In this paper we report a compact,
transportable laser system which is
suitable for such experiments.  In the future it can replace
stable but rather bulky dye-laser systems
\cite{Fischetal}.

\begin{figure}
\begin{center}
\includegraphics[width=0.45\textwidth]{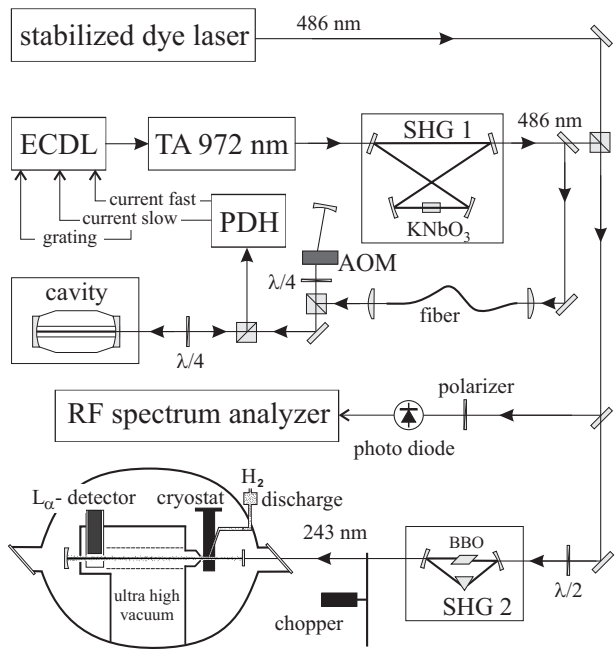}
\end{center}
\caption{A schematic diagram of the experimental setup. ECDL is the
external cavity diode laser at 972 nm, TA is the tapered
amplifier, SHG 1,~2 are the second harmonic generation stages,
PDH represents the Pound-Drever-Hall lock, and AOM is the acousto-optical
modulator used for scanning
the laser frequency. The output of the stabilized dye laser at 486 nm is
superimposed with the output of the SHG 1 and is used both for the
beat signal detection using a radio frequency (RF) spectrum analyzer
and for generation of 243 nm radiation.
The hydrogen spectra are detected alternatively either using
the dye laser system or the solid-state one. }
\label{fig1}
\end{figure}

The laser source is a frequency-quadrupled master-oscillator
power-amplifier (MOPA) system \cite{toptica} (see Fig. \ref{fig1}).
It is similar to the laser source demonstrated in Ref. \cite{zimmermann95}.
An anti-reflection coated laser diode operating at 972 nm is placed
in an external Littrow cavity.  The output is amplified using
a tapered amplifier (TA),
producing up to 650 mW at 972 nm.  This is coupled to the
first doubling stage (SHG 1 in Fig. \ref{fig1})
consisting of a bow-tie build-up cavity around a KNbO$_3$
crystal.  The crystal is used at normal incidence
and is cut for phase matching
at a temperature of 30 degrees.
The cavity is locked to
the laser using the Pound-Drever-Hall (PDH)
technique \cite{PDH}, with a modulation frequency of 20 MHz.
We have measured up to 210 mW at 486 nm.  The astigmatism in this
beam profile is corrected using a telescope
with cylindrical lenses.

The light at 486 nm is split into two beams.
 About 10 mW is taken for frequency stabilization of the laser,
 while the remainder is sent to the second doubling stage (SHG 2 in
 Fig. \ref{fig1}) after
 proper  mode-matching. The second frequency-doubling stage is
 a compact delta shape build-up cavity around a
 BBO crystal \cite{spectra}.
 After astigmatism compensation, we measure up to 20 mW at 243
 nm in a near-Gaussian profile.
 The master oscillator, tapered amplifier, both doubling
 stages (SHG 1 and SHG 2), and some auxiliary optics are mounted on a single
 $60 \times120$\ cm$^2$ optical breadboard.

 After a few months of operation, the output of the system
 decreased due to degradation of the TA beam profile.
 Laser powers were limited to  130 mW at 486 nm and 4.5 mW at 243 nm.
 Measurements reported in this paper were performed at
 this power level, which was enough to detect the $1S-2S$ transition.

The laser is actively stabilized to a high-finesse cavity at 486
nm \cite{Claudia}. Cavity mirrors are optically contacted to a
spacer produced from ultra-low expansion (ULE) glass.
The finesse of the cavity is 75,000, and the transmission peak
is about 15 kHz full width at half maximum (FWHM). The cavity rests
on a ceramic support in a small vacuum chamber and can be fixed for
transportation. The chamber is surrounded by single-stage temperature
stabilization system and is placed in a $60\times60\times60$\
cm$^3$ isolating box.

The ULE cavity is placed on a separate optical table.  The 486 nm
light from SHG 1 is delivered to the cavity by an
optical fiber. To compensate for the frequency noise introduced by
the fiber, a standard fiber noise compensation scheme is
implemented. The laser frequency is stabilized to the cavity by a
PDH lock at 11 MHz. The laser frequency can be shifted in respect
to the frequency of the cavity mode using a double-pass
acousto-optical modulator. The fact that we use the second
harmonic for generating the PDH error signal does not change the
phase characteristics of the lock since the transmission peak of
the SHG 1 is much broader then the bandwidth of the laser
frequency noise.

The error signal is split into three channels.  Fast current feedback
is sent directly
to the laser diode.  Slow feedback is sent to the
modulation input of the current
controller. The low-frequency drift of the laser cavity is also
compensated, using slow feedback to the grating.
This feedback configuration is similar to that used in 
Refs. \onlinecite{laserstab} and
\onlinecite{bianchini}.  The amplitude of the closed-loop
error signal measured in 50 kHz bandwidth  corresponds to 
frequency noise on the level of 100 Hz at 486 nm.

\begin{figure}
\begin{center}
\includegraphics[width=0.45\textwidth]{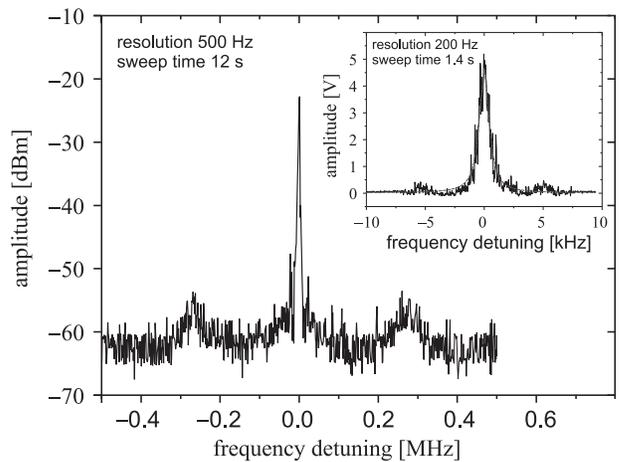}
\end{center}
\caption{ Power spectrum of the beat note between the second
harmonic of the diode laser and the dye laser in a logarithmic
scale. Inset: The zoomed central part of the spectrum
plotted on a
linear scale. The dashed line shows a Lorentzinan fit
with a 1 kHz FWHM. } \label{fig2}
\end{figure}

The power spectrum of the diode laser is studied by help of the
independent stabilized dye laser used for the previous
measurements \cite{Fischetal}. The dye laser is locked to a
non-transportable high-finesse ULE cavity which has significantly
better acoustic isolation and temperature control compared to the
cavity used for the diode laser stabilization. The typical
frequency drift of the stabilized dye laser is on the level of
about 0.2 Hz/s and its spectral line width
$\Delta\nu_{\rm{dye}}$ is characterized as 60 Hz at
486 nm \cite{BadHonnef}.
The power spectrum of the beat note between the two laser systems is
presented in Fig.\ref{fig2}. The spectrum is defined by
characteristics of the diode laser system, since on the given scale
the dye laser possesses a negligible spectral line width and
frequency drift and is essentially free of sidebands.

One can observe  broad sidebands
originating from the fast current feedback of the diode laser. The
amplitude of the sidebands is sensitive to the strength of the feedback:
with stronger feedback, more laser power is pushed into
the sidebands. The central peak of the spectrum has a width of
1.0~kHz FWHM at 486 nm. This mainly stems
from acoustic vibrations of the transportable cavity to which
the diode laser is locked.

The frequency drift of the diode laser can also
be characterized  using
beat-note measurements. It is approximately
10 Hz/s at 486 nm, more then ten times higher then the
frequency drift of the non-transportable cavity. Once
again, this difference is attributable mainly to the 
optical cavity to which the diode laser is locked.
The transportable cavity uses a simpler temperature 
controller and has greater thermal coupling to its 
ceramic support.

\begin{figure}
\begin{center}
\includegraphics[width=0.45\textwidth]{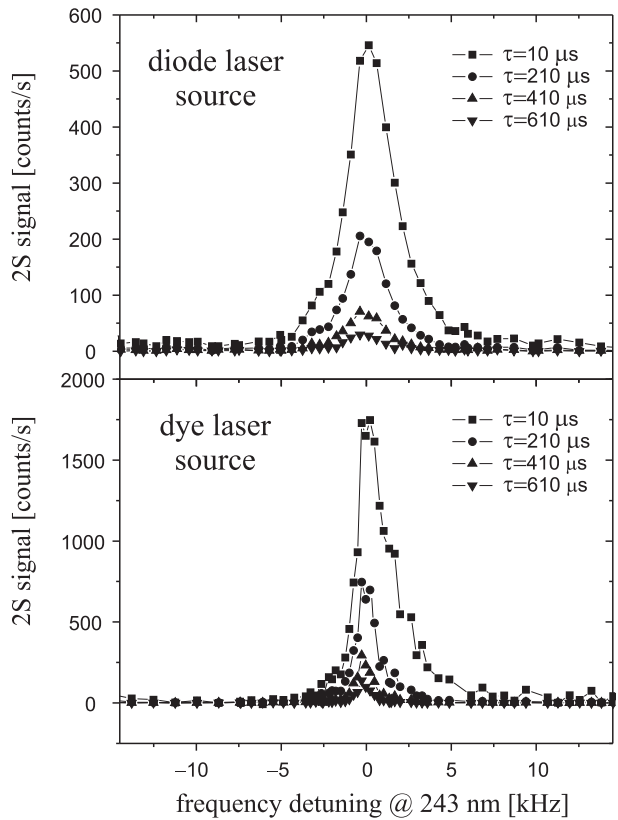}
\end{center}
\caption{ Top: $1S-2S$ time-resolved spectra recorded with the diode
laser source. The power of 243 nm radiation coupled to the enhancement
cavity equals 2.6 mW. Bottom: $1S-2S$ spectra recorded with the dye
laser source at the power of 2.2 mW.  In each case two spectra
recorded in two different scan directions are averaged and the
drift of the corresponding reference cavity is compensated.}
\label{fig3}
\end{figure}

We tested the new diode laser source using the hydrogen spectrometer
\cite{Fischetal, BadHonnef} depicted schematically  in
Fig.\ref{fig1}. The 243 nm radiation is coupled to a linear
enhancement cavity, where the
excitation of the hydrogen beam takes place. The flow of atomic
hydrogen produced in a radio-frequency gas discharge is cooled to
5 K by a flow-through cryostat. After a 3 ms excitation phase, the
243 light is blocked by a chopper operating at 160 Hz, and the
detection phase begins. Atoms in the $2S$ state coming to the
detection zone are quenched in a weak electric field.  Emitted
Lyman-alpha photons are detected by a photomultiplier tube, and
counts are accumulated until the beginning of a new excitation phase.
By introducing a delay time $\tau$ between the end of the excitation
and the start of detection, we select slow atoms from the original
Maxwellian distribution. The delay of $\tau= 1$ ms corresponds to
the cut-off velocity of 130 m/s. For each laser frequency,
the measurement cycle is repeated for 1 s, and photon counts 
are accumulated by a multi-channel scaler. Thus we
simultaneously record a number of delayed spectra corresponding to
different velocity classes. This detection scheme allows for
evaluation of the second-order Doppler effect and reduces the
background level that is due to scattered 243 nm radiation.

A time-resolved spectrum of the $1S-2S$ transition excited using 
the new laser system is presented in Fig.\ref{fig3} (top), where the
photon count rate is plotted versus frequency detuning at 243 nm.
The spectral line width of the transition $\Delta f^{1S-2S}_{\rm
{diode}}$ equals 2.8(1) kHz FWHM for $\tau=10\ \mu$s and reduces
to 2.2(1) kHz for $\tau=610\ \mu$s. Besides the contribution
of the finite line width of the excitation radiation, the $1S-2S$
transition line width is defined by time-of-flight broadening,
power broadening, ionization broadening, and the second-order
Doppler effect. The delayed detection significantly reduces the
time-of-flight broadening and the second-order Doppler effect due to
the velocity selection. To analyze the spectrum, we compare it to
the calculated line shape in the case of an infinitely narrow-band
excitation spectrum using a Monte-Carlo approach
\cite{MonteCarlo}. For the delay $\tau=610\ \mu$s, the simulated
line has a FWHM at 243 nm of $\Delta f_{\rm theor}(610\ \mu\rm{s})=0.62$ kHz
when 2.6 mW of  243 nm radiation coupled to
the enhancement cavity.

Assuming Lorentzian profiles, one can write the following
approximate relation:
\begin{equation}\label{widths}
\Delta f^{1S-2S}_{\rm {diode}}(\tau)\simeq \Delta f_{\rm
theor}(\tau)+ 2\,\Delta\nu_{\rm{diode}},
\end{equation}
where $\Delta\nu_{\rm{diode}}=2.0$~kHz is the measured spectral
width of the diode laser system at 243 nm. Factor 2 results from
two-photon excitation of the $1S-2S$ transition. In this
particular case, the line width of the $1S-2S$ transition for longer
delay times is mainly defined by the spectral line width of the
excitation radiation. The spectral resolution of the spectrometer
$f_0/\Delta f^{1S-2S}_{\rm {diode}}(610\ \mu\rm{s})$ equals
$6\times 10^{11}$.

To compare the diode laser source with the dye laser, we also performed a set
of measurements with the dye laser coupled to the same doubling stage
as depicted in  Fig.\ref{fig1}. A
$1S-2S$ spectrum recorded using the dye laser at a similar excitation
power level is shown in Fig.\ref{fig3}(bottom).
The recorded lines are considerably narrower compared
to the case of the diode laser.  The line shape asymmetry resulting
from the second-order Doppler effect is clearly visible for short
delays. The amplitude noise partly results from the intensity
fluctuations of the dye laser. The widths of the delayed
spectra are close to the calculated ones: for example, $\Delta
f^{1S-2S}_{\rm {dye}}(610\ \mu\rm{s})=0.75(5)$~kHz at 243 nm.

The $2S$ count rate using the diode laser is approximately three
times lower than using the dye laser. Besides slightly  different
excitation powers, the difference in the count rates can be
explained by two factors: (i)  the broader line width of the diode
laser and (ii) the distribution of the laser diode power over a
wide spectral  range. The loss of the $2S$ counts in the narrow
$1S-2S$ line is caused by distributed features in the spectrum of
the diode laser.

Since the dye laser spectrum is practically free from the
background, one can evaluate separately the effect of the
distributed spectrum of the diode laser. To do it, we have to take
into account the difference in the spectral widths of the central
peak of the diode laser spectrum used for spectroscopy and the
spectral width of the dye laser. Under the realistic
approximations $\Delta f_{\rm theor}\gg 2 \Delta \nu_{\rm{dye}}$
and $\Delta f_{\rm theor}\ll 2 \Delta \nu_{\rm{diode}}$, it is
equivalent to the comparison of the areas under corresponding
transitions normalized to the 243 nm power squared.
The $2S$ excitation efficiency for  the diode laser source is
$40(10)$\% of the efficiency for the dye laser.  This number is
constant for all available delays. This means that for the same
excitation powers and for the same line widths of the lasers, the
diode laser would excite only 40\% of the atoms in a narrow
spectral line compared to the dye laser.

It should be possible to narrow the spectrum of the diode laser
using a two-stage, two-cavity lock scheme.  Our diode laser
is stabilized to a high-finesse cavity, and has a spectral
width of approximately 1 kHz at 486 nm.  This residual width
could be removed relative to a second cavity using an AOM.
It is not enough simply to increase the feedback loop gain
in the present setup.  Increasing the gain only produces
larger sidebands in the beat signal spectrum (see Fig. \ref{fig2})
and a subsequent loss of $2S$ excitation efficiency.

In summary, a high-power narrow-bandwith diode laser source
is reported that is capable of driving the $1S-2S$ transition
in atomic hydrogen 
with a high signal-to-noise ratio.  The spectral width of the
laser is 2 kHz at 243 nm, limited mainly by the characteristics
of the small, transportable optical cavity.
Implementing the
new concept of vertically-suspended reference cavities
\cite{Notcutt} and a spacer with the proper choice of
ULE zero-expansion point
will significantly increase the short- and long-term frequency
stability of the diode laser source while keeping the whole system
compact and transportable. This laser system is a promising source for
spectroscopy of exotic hydrogen-like systems such as
anti-hydrogen, tritium, positronium, and muonium.

N.K. acknowledges the support of DFG (grant No.
436RUS113/769/0-1) and RFBR (grants. No. 03-02-04029, 04-0217443).  N.K.
and S.D.B. acknowledge the support of the Alexander von Humboldt Foundation.

\end{document}